\documentclass[11pt]{article}
\usepackage{moriond,epsfig}
\usepackage[latin1]{inputenc}
\usepackage[OT1]{fontenc}
\usepackage{latexsym}
\def\lsim{\raisebox{-.4ex}{$\stackrel{<}{\scriptstyle \sim}$\,}}

\def\be{\begin{equation}}
\def\ee{\end{equation}}
\def\bea{\begin{eqnarray}}
\def\eea{\end{eqnarray}}

\begin{document}
\vspace*{0.9cm}
\title{VECTOR MODELS FOR DARK ENERGY}

\author{Jose Beltr\'an Jim\'enez and  Antonio L. Maroto}

\address{Departamento de F\'{\i}sica Te\'orica I, 
Universidad Complutense de Madrid,\\
28040 Madrid, Spain}

\maketitle\abstracts{
We explore the possibility that the present stage of accelerated
expansion of the universe is due to the presence of a cosmic
vector field. We show that vector theories allow for the generation
of an accelerated phase without the introduction of potential terms or
unnatural scales in the Lagrangian. We propose a particular model
with the same number of parameters as $\Lambda$CDM and excellent fits
to SNIa data. The model is scaling during radiation era, with natural
initial conditions, thus avoiding the cosmic coincidence problem.  
Upcoming observations will be able  
 to clearly discriminate it from standard $\Lambda$CDM cosmology.}

%\section{Introduction}
The fact that today the dark energy density is comparable to the 
matter energy density poses one of the most important problems in order to
find viable 
models of dark energy. Indeed, to achieve this, most of the
models, not only the cosmological constant, but also 
those based on scalar fields such as quintessence or k-essence, and
 modified gravity theories such as $f(R)$, DGP, etc, require
the  introduction of unnatural scales either in their Lagrangians or 
in their initial conditions. This is the so called 
{\it cosmic coincidence problem} (see \cite{review} and references therein).

Therefore, we would like to find a model without  dimensional scales 
(apart from Newton's constant $G$), with the same number of parameters as 
$\Lambda$CDM, with natural initial conditions
 and with good fits to observations. In addition, the model should be stable
under small perturbations. We will show that vector models can do the
job \cite{BM}.

%\section{A vector-tensor model}

Let us start by writing the action of our 
vector-tensor theory of gravity containing only 
two fields and two derivatives and without potential terms 
(see  \cite{potential} for previous works on vector models for 
dark energy with potential terms):
\begin{eqnarray}
S=\int d^4x \sqrt{-g}\left(-\frac{R}{16\pi G}
-\frac{1}{2}\nabla_\mu A_\nu\nabla^\mu A^\nu
+\frac{1}{2} R_{\mu\nu}A^\mu A^\nu\right)
\label{action}
\end{eqnarray}
Notice that the theory contains no free parameters, the only dimensional
scale being the Newton's constant.   
The numerical factor
in front of the vector kinetic terms can be fixed 
by the field normalization.
Also notice that  
 $R_{\mu\nu}A^\mu A^\nu$ can be written as a combination of derivative terms as 
$\nabla_\mu A^\mu\nabla_\nu A^\nu-\nabla_\mu A^\nu\nabla_\nu A^\mu$.

The classical equations of motion derived from the action in (\ref{action})
are the Einstein's and
 vector field equations:
\begin{eqnarray}
R_{\mu\nu}-\frac{1}{2}R g_{\mu\nu}&=&8\pi G (T_{\mu\nu}+T_{\mu\nu}^A) \label{eqE}\\
\Box A_\mu + R_{\mu\nu}A^\nu&=&0 \label{eqA}
\end{eqnarray}
where $T_{\mu\nu}$ is the conserved energy-momentum tensor for matter and radiation and
$T_{\mu\nu}^A$ is the energy-momentum tensor coming from the vector field.
For the simplest isotropic and
 homogeneous flat cosmologies,  we assume that 
the spatial components of the vector field vanish, so that 
$A_\mu=(A_0(t),0,0,0)$ and that the 
space-time geometry will 
be given by:
\begin{equation}
ds^2=dt^2-a^2(t)\delta_{ij}dx^idx^j,
\label{metric}
\end{equation}
For this metric (\ref{eqA}) reads:
\begin{eqnarray}
\ddot{A}_0+3H\dot{A}_0-3\left[2H^2+\dot{H}\right]A_0=0\label{fieldeq0}
\end{eqnarray}
Assuming that the universe has gone through  radiation and 
matter phases in which
the contribution from dark energy was negligible, we can easily solve
this equation in those periods.  In that case, the
above equation has a growing and a decaying solution:
\begin{eqnarray}
A_0(t)=A_0^+t^{\alpha_+}+A_0^-t^{\alpha_-}\label{fieldsol}
\end{eqnarray}
with $A_{0}^\pm$ constants of integration and $\alpha_{\pm}=-(1\pm 1)/4$
 in the radiation era, and
 $\alpha_{\pm}=(-3\pm\sqrt{33})/6$ in the matter era. 
On the other hand, the $(00)$ component of Einstein's equations reads:
\begin{eqnarray}
H^2=\frac{8\pi G}{3}
\left[\sum_{\alpha=M,R} \rho_\alpha+\rho_{A}\right]
\label{Friedmann}\end{eqnarray}
where the vector energy density is given by: 
\begin{eqnarray}
\rho_{A}=\frac{3}{2}H^2A_0^2+3HA_0\dot A_0-\frac{1}{2}\dot A_0^2
\end{eqnarray}
\begin{figure}
\vspace{0.3cm}
\begin{center}{\epsfxsize=7.5 cm\epsfbox{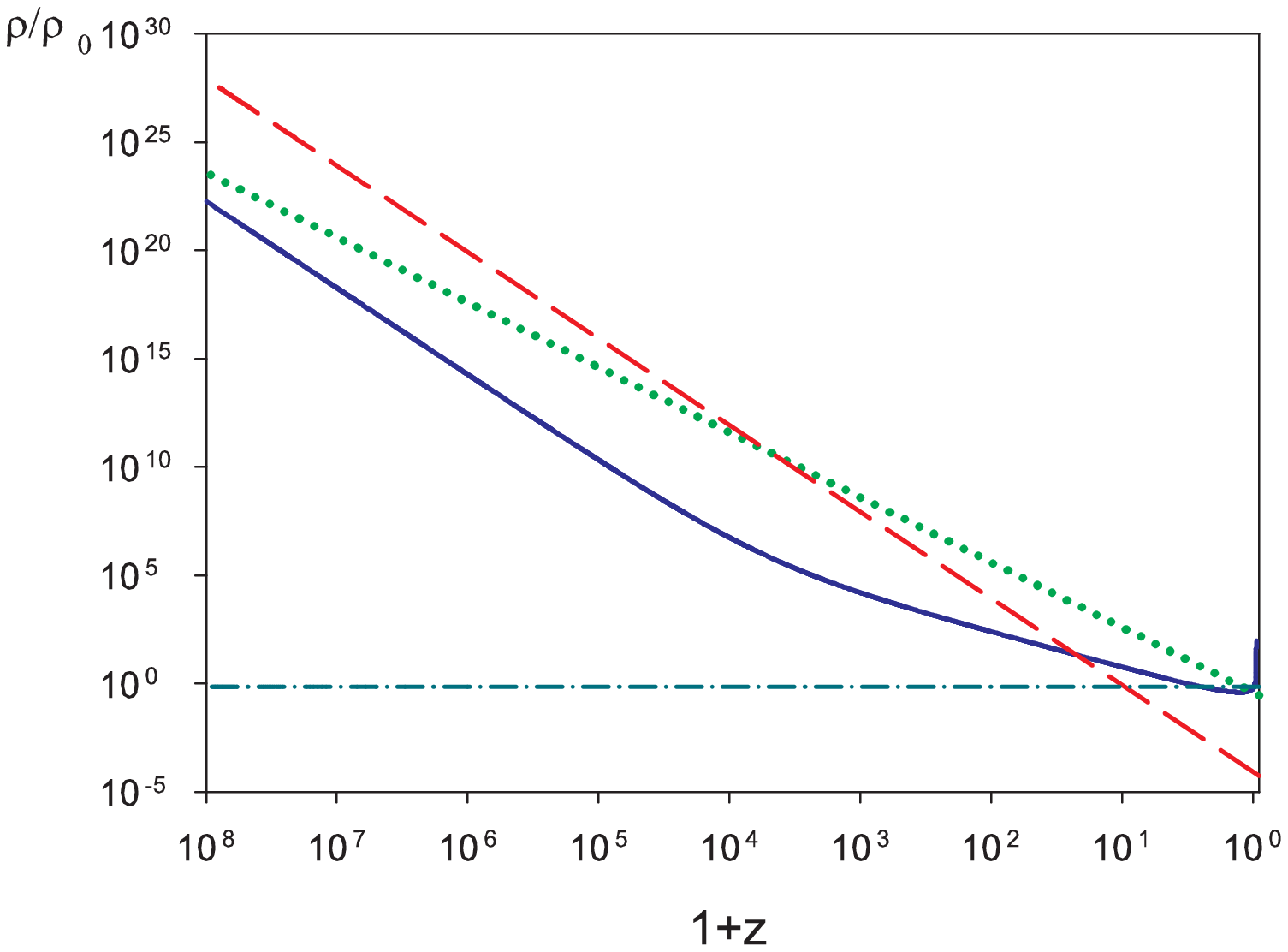}}\hspace{1.0cm}
{\epsfxsize=7 cm\epsfbox{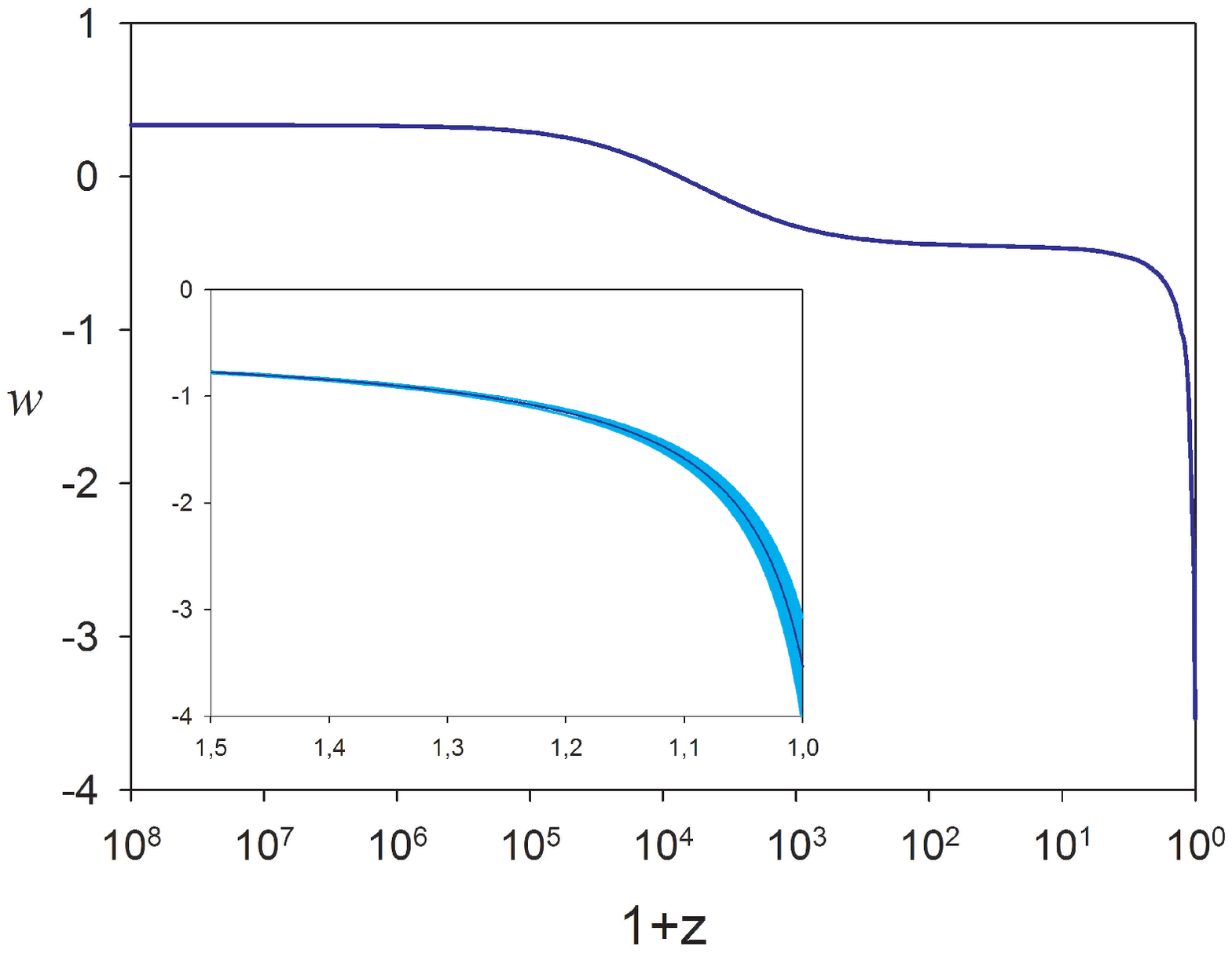}}
\caption{\footnotesize (Left) Evolution of energy densities for the best fit model. 
Dashed (red) for
radiation, dotted (green) for matter and solid (blue) for vector 
dark energy. We show also for comparison the cosmological constant
density in dashed-dotted line.
(Right) Evolution of dark energy equation of state for
the best fit model. The lower panel shows the 1$\sigma$ confidence interval.}
\end{center}
\end{figure}
Using the
growing mode solution from (\ref{fieldsol}), we obtain 
$\rho_{A}= \rho_{A0} a^\kappa$ 
with $\kappa=-4$ in the radiation era and 
$\kappa=(\sqrt{33}-9)/2 \simeq -1.63$ in 
the matter era. Thus, the energy density of the vector field 
starts scaling as radiation
 at early times, so that $\rho_A/\rho_R=$ const.  
However, when the universe enters its matter era, $\rho_A$ starts growing 
relative to 
$\rho_M$ eventually overcoming it at some point, in which the dark energy 
vector  
field would become
the dominant component (see Fig. 1).  Notice that  
since $A_0$ is essentially constant during 
radiation era, solutions do not depend on the precise initial time
at which we specify it.  Thus, 
once the present value of the Hubble parameter $H_0$
and the constant $A_0$ during radiation (which fixes the total matter
density $\Omega_M$) are specified, 
the model is completely 
determined, i.e. this model contains the same
number of parameters as $\Lambda$CDM, which is the minimum number 
of parameters
of a cosmological model with dark energy. As seen from
Fig.1  the evolution of the universe ends at a finite time $t_{end}$ with a 
 singularity in which
$a\rightarrow a_{end}$ with $a_{end}$ finite, 
$A_0(t_{end})=M_P/(4\sqrt{\pi})$,  
$\rho_{DE}\rightarrow \infty$ and
$p_{DE}\rightarrow -\infty$. 
 
We can also calculate the effective equation of state for dark energy
as:
\begin{eqnarray}
w_{DE}=\frac{p_A}{\rho_A}=\frac{-3\left(\frac{5}{2}H^2+\frac{4}{3}\dot{H}\right)A_0^2+HA_0\dot{A}_0
-\frac{3}{2}\dot{A}_0^2}{\frac{3}{2}H^2A_0^2+3HA_0\dot A_0-\frac{1}{2}\dot A_0^2}
\end{eqnarray}
Again, using the approximate solutions in (\ref{fieldsol}), we obtain:
$w_{DE}=1/3$ in the radiation era and $w_{DE}\simeq -0.457$ in the 
matter era. As shown in Fig. 1, the equation of state can cross the so called phantom 
divide, so that we can have $w_{DE}(z=0)<-1$.

In order to confront the  predictions of the model with observations
of high-redshift supernovae type Ia, we have carried out  a
$\chi^2$ statistical analysis for two  supernovae datasets, namely, 
the Gold set \cite{Gold}, 
containing 157 points with $z < 1.7$, and  
the more recent SNLS  data set \cite{SNLS}, comprising 115 
supernovae but with lower redshifts ($z < 1$). In Table 1 we 
show the results for the best fit together 
with its corresponding $1\sigma$  intervals for the 
two data sets. We also
show for comparison the results for a standard
$\Lambda$CDM model. We see that the vector model (VCDM) fits the data
considerably
better than  $\Lambda$CDM (in more than $2\sigma$)  in the Gold 
set, whereas the situation is reversed in the SNLS set. This is just 
a reflection of the well-known $2\sigma$ tension \cite{tension}
between the two 
data sets.  Compared with $\Lambda$CDM, we see that VCDM
favors a younger universe (in $H_0^{-1}$ units)
with larger matter density. In addition, 
the deceleration-aceleration transition takes place at a lower redshift
in the VCDM case. The present value
of the equation of state  with $w_0=-3.53^{+0.46}_{-0.57}$ 
which clearly excludes the cosmological constant value $-1$. Future
surveys \cite{Trotta} are expected to be able to measure $w_0$ at the 
few percent level
and therefore could discriminate between the two models. 
\begin{table}
\begin{center}
\footnotesize{\begin{tabular}{|c|c|c|c|c|}
\hline
  & & & &\\ 
 & VCDM  & $\Lambda$CDM  & VCDM & $\Lambda$CDM \\
& \footnotesize{Gold}  &\footnotesize{Gold} & 
\footnotesize{SNLS}   & \footnotesize{SNLS} \\
& & & &\\
\hline & & & & \\$\Omega_M$ &$0.388^{+0.023}_{-0.024}$ & $0.309^{+0.039}_{-0.037}$ & 
$0.388^{+0.022}_{-0.020}$ & $0.263^{+0.038}_{-0.036}$
\\& & && \\ 
\hline & & & &\\ $w_0$ &$-3.53^{+0.46}_{-0.57}$& $-1$ & $-3.53^{+0.44}_{-0.48}$ & $-1$
\\& & & & \\
\hline  & & & &\\$A_0$ & $3.71^{+0.022}_{-0.026}$ &--- & $3.71^{+0.020}_{-0.024}$&---
\\\footnotesize{$(10^{-4}\;M_P)$}& & & & \\
\hline  & & & &\\$z_T$ &$0.265^{+0.011}_{-0.012}$ &$0.648^{+0.101}_{-0.095}$ & $0.265^{+0.010}_{-0.012}$&$0.776^{+0.120}_{-0.108}$ 
\\& & & & \\
\hline  & & & &\\$t_0$   &$0.926^{+0.026}_{-0.023}$ & $0.956^{+0.035}_{-0.032}$& 
$0.926^{+0.022}_{-0.022}$ &$1.000^{+0.041}_{-0.037}$ 
\\\footnotesize{$(H_0^{-1})$}& &  & & \\
\hline & & & &\\$t_{end}$ &$0.976^{+0.018}_{-0.014}$ & --- &$0.976^{+0.015}_{-0.013}$ & ---
 \\\footnotesize{$(H_0^{-1})$}& & & &\\
\hline & & & &\\ $\chi^2_{min}$ & 172.9& 177.1 &115.8 & 111.0
\\ & & & &\\ 
\hline
\end{tabular}}
\caption{\footnotesize{Best fit parameters with $1\sigma$ 
intervals for the vector model (VCDM) and
the cosmological constant model ($\Lambda$CDM) for the Gold (157 SNe)
and SNLS (115 SNe) data sets. $w_0$
denotes the present equation of state of dark energy. $A_0$ is 
the constant value of the vector field component during radiation. 
$z_T$ is the deceleration-aceleration transition redshift.
$t_0$ is the age of the universe in units of the present
Hubble time. $t_{end}$ is the duration of the universe in 
the same units.}}
\end{center}
\end{table} 
We have also compared with other parametrizations
for the dark energy equation of state \cite{models}. Since our 
one-parameter fit has a reduced chi-squared: 
$\chi^2/d.o.f=1.108$,  VCDM provides 
the best fit to date for the Gold data set. 

We see that unlike the 
cosmological constant case, throughout radiation era
 $\rho_{DE}/\rho_R\sim 10^{-6}$ in our case. Moreover the scale of the vector
field $A_0=3.71 \times 10^{-4}$ $M_P$ in that era is 
relatively close to the Planck scale and could arise naturally  
in the early universe without the need of introducing extremely small 
parameters.

In order to study the model stability we have considered the evolution 
of metric and vector field perturbations. Thus, we obtain the dispersion
relation and  the propagation speed of  scalar, vector and tensor modes.  
For all of them we obtain $v=(1-16\pi G A_0^2)^{-1/2}$ which is
 real throughout the universe evolution, since the value $A_0^2=(16\pi G)^{-1}$
exactly corresponds  to that at the  final singularity. Therefore
 the model does not exhibit exponential instabilities. As shown in
\cite{Mukhanov}, the fact that the propagation speed is faster than $c$ 
does not necessarily implies inconsistencies with causality. We have 
also considered the evolution of scalar perturbations in the vector field 
generated by scalar metric perturbations during matter and radiation eras, 
and found that the energy density contrast
$\delta \rho_A/\rho_A$ is constant on super-Hubble
scales, whereas it oscillates with growing amplitude as $a^2$ 
in the radiation  era and as $\sim a^{0.3}$ in the matter era for 
sub-Hubble scales.  
Therefore  again, we do not find exponentially 
growing modes.

If we are interested in extending the applicability range of the model
down to solar system scales then we should study the 
corresponding post-Newtonian parameters (PPN). We can see that
 for the model in (\ref{action}), the static PPN parameters
 agree with those of 
General Relativity \cite{Will}, i.e. 
$\gamma=\beta=1$. 
For the parameters  
associated to preferred frame effects we get: $\alpha_1=0$ and
$\alpha_2=8\pi  A_\odot^2/M_P^2$ where $A_\odot^2$ is the norm of the 
vector field at the solar system scale. Current limits 
$\alpha_2 \;\lsim 10^{-4}$ (or $\alpha_2\;\lsim 10^{-7}$ for static vector 
fields during
solar system formation) then impose a bound $A_\odot^2\lsim \, 
10^{-5}(10^{-8})
\,M_P^2$. In order to determine whether such bounds conflict with
the model predictions or not,  we should know the predicted value of 
the field at solar system scales, which in principle does not need to 
agree with the
cosmological value. 
Indeed, $A_\odot^2$  will be determined by the mechanism
that generated this field in the early universe characterized by its 
primordial spectrum of perturbations, and the
subsequent evolution in the formation of the galaxy and solar system. 
 Another potential difficulty arising generically in
 vector-tensor models is the presence of negative energy modes for
perturbations 
on sub-Hubble scales. They are known to lead to instabilities at the 
quantum level,  but not necessarily at the classical level as we have
shown previously. 
In any case, the model proposed is not intended as
a quantum theory of the gravitational interaction, which would be
beyond the scope of this work.

 In conclusion,  vector theories offer an accurate 
phenomenological description of dark energy in which fine tuning problems
 could be easily avoided.

{\em Acknowledgments:} This work has been  supported by
DGICYT (Spain) project numbers FPA 2004-02602 and FPA
2005-02327, UCM-Santander PR34/07-15875 and by CAM/UCM 910309. 
J.B. aknowledges support from MEC grant
BES-2006-12059.
%\vspace*{-0.2cm}
\section*{References}
%\vspace*{-0.2cm}

\end{document}